\documentclass[prl,showpacs,amsmath,amssymb,superscriptaddress,letterpaper,preprintnumbers]{revtex4}

\usepackage{graphicx}
\usepackage{amssymb,amsmath}
\usepackage{amsthm}
\usepackage{bm}
\DeclareMathOperator{\M}{\mathcal{M}}
\DeclareMathOperator{\X}{\mathcal{X}}

\begin{document}

\preprint{APS/123-QED}

\title{Neural Population Coding is Optimized by Discrete Tuning Curves}

\author{Alexander P. Nikitin}
\email{a.nikitin@warwick.ac.uk}
\affiliation{School of Engineering, University of Warwick, Coventry CV4 7AL, UK}

\author{Nigel G. Stocks}
\email{n.g.stocks@warwick.ac.uk}
\affiliation{School of Engineering, University of Warwick, Coventry CV4 7AL, UK}

\author{Robert P. Morse}
\email{r.p.morse@aston.ac.uk}
\affiliation{School of Life and Health Sciences, Aston University, Birmingham B4 7ET, UK}

\author{Mark D. McDonnell}
\email{mark.mcdonnell@unisa.edu.au}
\affiliation{Institute for Telecommunications Research, University of South Australia, SA 5095, Australia}

\date{\today}

\begin{abstract}
The sigmoidal tuning curve that maximizes the mutual information for a Poisson neuron, or population of Poisson neurons, is obtained. The optimal tuning curve is found to have a discrete structure that results in a quantization of the input signal. The number of quantization levels undergoes a hierarchy of phase transitions as the length of the coding window is varied. We postulate, using the mammalian auditory system as an example, that the presence of a subpopulation structure within a neural population is consistent with an optimal neural code.
\end{abstract}



\maketitle

Neuronal responses often appear noisy in the sense that repeated presentation of identical stimuli result in variable action potential timings. This variability is often closely modeled by Poisson statistics~\cite{NeuronModels,spikes} and, hence, the Poisson neuron has become an archetypal model for neural rate coding. In this model the input signal $x$ is coded in the mean firing rate $\nu=g(x)$ where $g(x)$ is known as the {\em tuning curve} (or the stimulus-response curve, gain function or rate-level function). While several definitions of {\em rate} exist~\cite{Lansky.04}, following related studies~\cite{Brunel.98,Bethge1,Bethge2}, here we assume the {\em observable} output when the mean rate is $\nu$ is the number of spikes, $k$, that occur in a time window $T$. The input $x$ is assumed to be a continuous variable, such as an external sensory stimulus.

Despite the popularity of the Poisson neural model, remarkably the $g(x)$ that maximizes Shannon mutual information~\cite{Blahut,Quiroga.09} has not been obtained, except in the limit $T \rightarrow \infty$ ~\cite{Brunel.98,McDonnell.PRL08}. Arguably, this limit is not relevant to a large number of biological sensory systems where it is well established that behavioral responses occur on timescales that imply short coding windows~\cite{spikes}. In this letter we obtain the optimal tuning curve for finite $T$.

Our main finding is that the optimal tuning curve is discrete, in the sense that many stimuli values result in the same mean firing rate. The number of discrete levels, $M$, increases as $T$ increases. This result means that when mutual information is to be maximized, signal quantization is an emergent feature of the optimal coding scheme and is superior to analogue coding. We also demonstrate this means neural {\em subpopulations} might be necessary to optimize an overall population.

This result of optimal $M$-ary coding differs significantly
from~\cite{Bethge1,Bethge2}, which predicts a single phase transition from
binary to continuous tuning curves---i.e. from discrete to analog coding. This difference is because we maximize mutual information, while~\cite{Bethge1,Bethge2} minimizes mean square error (MSE). We consider mutual information, rather than other metrics like MSE~\cite{Bethge1} and Fisher information~\cite{Bethge.02}, because it does not rely on assumptions about how a neuron's response may be `decoded'~\cite{Quiroga.09}. Furthermore, mutual information is intimately linked to MSE via rate-distortion theory~\cite{Rose}.

To derive the optimal tuning curve, we make use of a known result from the photonics literature. The properties of Poisson neurons are very similar to those of direct detection photon channels, which operate by modulating the intensity of a photon emitting source. Both can be modeled as Poisson point processes~\cite{Teich.76}.

A classical problem in communication theory is that of finding the signal distribution that maximizes the mutual information for a channel. The resultant {\em optimal code}is said to achieve {\em channel capacity}~\cite{Blahut}. The optimal input distribution for the direct detection photon channel has been proven to be discrete~\cite{Kabanov,Shamai90}. Indeed, the discreteness of optimal input distributions is the norm, regardless of whether the output distribution is discrete or continuous~\cite{Smith71,Huang.05}.  Consequently, although we have assumed a rate code based on the discrete random variable defined by counting spikes (alternatively we could have defined it in terms of a continuous random variable based on interspike intervals), the central result in this paper is not dependent on the definition of rate but is rather a property of the `channel noise'---see~\cite{Huang.05,Ikeda.09}.

The discreteness of the optimal signal for the optical Poisson channel implies that the optimal stimulus for a Poisson neural system is also discrete. However, this is not physically realistic, as the distribution of an external stimulus is not controlled by a neural system, and is likely to be continuous, e.g.  speech, or natural sound statistics. Instead, it is plausible that a neural system may have been optimized by evolution so that the {\em tuning curve} discretizes its input to match the theoretical optimal source distribution.

The mutual information~\cite{Blahut,Quiroga.09} between the (continuous) input random variable $x\in\X=[x_{\min},x_{\max}]$, and the (discrete) output random variable, $k$, is
\begin{equation}
\label{m_information}
I(x;k)=\sum_{k=0}^{\infty}\int_{x\in\X}dx P_{x}(x) P[k|x] \log_{2}
\frac{P[k|x]}{P_{k}(k)},
\end{equation}
where
$P_{k}(k)=\int_{x\in\X}dx P_{x}(x) P[k|x]$. Here $P_{x}(x)$, $P_{k}(k)$ and
$P[k|x]$ are the distributions of the stimulus, the response and the
conditional distribution respectively.

For Poisson statistics, the conditional distribution is
\begin{equation}
\label{Poisson}
Q[k|\nu]=\frac{[T\nu]^{k}}{k!}\exp(-T\nu), \quad k=0,..,\infty.
\end{equation}
The mean firing rate is restricted to $\nu\in[\nu_{\min},\nu_{\max}]$, where
the upper bound $\nu_{\max}$ is due to physiological limits (metabolic and refractory), and we set $\nu_{\min}=0$. Later, we use the notation $N=T\nu_{\max}$ to denote the {\em maximum} mean spike count.

The conversion of a signal follows the Markov chain
$x\rightarrow\nu\rightarrow k$. Note that $k$ is observable from a short duration, $T$, while $\nu$ is not. We refer to $x\rightarrow\nu$ and $\nu\rightarrow k$ as separate `subchannels.' To find the optimal channel, we maximize the mutual information
by variation of the distribution $P_{\nu}(\nu)$ for given $P_{x}(x)$ and $Q[k|\nu]$.
Since the distribution of $\nu$ is
$P_{\nu}(\nu)=\int_{x\in\X}dx P_{x}(x) \delta(\nu-g(x))$,
where $\delta(.)$ is the Dirac delta function,
variation of $P_{\nu}(\nu)$ means variation of the tuning curve $g(x)$.

We now present the following theorem:
{\em The mutual information in the neural channel,
$x\rightarrow\nu\rightarrow k$, is maximized when
the distribution $P_{\nu}(\nu)$ is discrete.}

{\em Remark.}
The neural channel forms a Markov chain for which the following equations are valid,
\begin{eqnarray}
P(x,\nu,k)&=&P_{k|\nu}[k|\nu] P[\nu|x] P_{x}(x), \\
P(x,\nu,k)&=&P_{k|\nu}[x|\nu] P[\nu|k] P_{k}(k).
\end{eqnarray}
We assume $P_{k|\nu,x}[k|\nu,x]=P_{k|\nu}[k|\nu]$ and
$P_{x|\nu,k}[x|\nu,k]=P_{x|\nu}[x|\nu]$ due to the definitions of the
subchannels as $\nu=g(x)$, i.e. $P_{x|\nu}[x|\nu]=\delta(\nu-g(x))$,
where $g(x)$ is a single-branched function, and $P_{k|\nu}[k|\nu]=Q[k|\nu]$.

\begin{proof}
First we prove that $I(x;k)= I(\nu;k)$. From Theorem 5.2.8 of~\cite{Blahut},
the mutual information between the variable $k$ and the pair $(\nu,x)$
can be written in two ways,
\begin{eqnarray}
I(k;(\nu,x))&=&I(k;\nu)+I(k;x|\nu), \\
I(k;(\nu,x))&=&I(k;x)+I(k;\nu|x),
\end{eqnarray}
where the {\em conditional} mutual information expressions are
\begin{eqnarray}
I(k;x|\nu)&=&
\iint_{\substack{x\in\X \\ \nu\in\M}} dx d\nu \sum_{k=0}^{\infty}P(x,\nu,k)
\log_{2} \frac{P_{x|\nu,k}[x|\nu,k]}{P_{x|\nu}[x|\nu]}, \nonumber \\
I(k;\nu|x)&=& \iint_{\substack{x\in\X \\ \nu\in\M}} dx d\nu
\sum_{k=0}^{\infty}P(x,\nu,k)
\log_{2} \frac{P_{\nu|x,k}[\nu|x,k]}{P_{\nu|x}[\nu|x]}. \nonumber
\end{eqnarray}
Since $P_{x|\nu,k}[x|\nu,k]=P_{x|\nu}[x|\nu]$, the conditional mutual
information $I(k;x|\nu)$ is zero. Next, note that the variable $\nu$
directly depends on the random variable $x$, $\nu=g(x)$ and hence
$P_{\nu|x,k}[\nu|x,k]=P_{\nu|x}[\nu|x]$
and the conditional mutual information $I(k;\nu|x)$ is also zero.
Consequently, we are left with the following two equations,
$I(k;(\nu,x))=I(k;\nu)$ and $I(k;(\nu,x))=I(k;x)$. This means that
$I(k;x)=I(k;\nu)$, and the mutual information in the neural channel
is equal to the mutual information of the noisy subchannel.

To proceed, we now consider the noisy neural subchannel $\nu\xrightarrow{Q[k|\nu]}k$, and use a theorem from~\cite{Shamai90} for a `direct detection' photon channels, where the input is a continuous time inhomogeneous Poisson rate, $\lambda(t)$. Due to `bandwidth constraints,' $\lambda(t)\le A$ is constant during equal durations $\Delta$. The output is the sequence of photon arrival times within $\Delta$, $\{t_i\}_{i=1}^y$ . A key result in~\cite[Section~3]{Shamai90} is that this channel is mathematically equivalent to one where the output is the photon count $y$ within $\Delta$, and the input is a time-independent variable $\lambda$. The latter channel's distribution is given by~\ref{Poisson}, except instead of counting $y$ photons in response to $\lambda\le A$ during $\Delta$, the neural Poisson channel output is the spike count, $k$ in response to $\nu\le\nu_{\max}$ during $T$.  Thus, $I(\lambda(t);\{t_i\}_{i=1}^y)=I(\lambda;y)\equiv I(\nu;k)=I(k;x)$, and since \cite[Theorem~1]{Shamai90} states that the optimal distribution of $\lambda$ is discrete, the mutual information in the neural channel is also maximized when $P_\nu(\nu)$ is discrete.\qedhere
\end{proof}

The proven theorem does not provide any means for finding a closed-form solution for the optimal discrete distribution, $P_\nu(\nu)$. However, its utility is that it allows a reduction in the set of functions we need to consider when optimizing $P_{\nu}(\nu)$ and/or the tuning curve
$g(x)$.

Without loss of generality we can now introduce the following simplifying restriction
 for the function $g(x)$. Let $g(x)$ be a non-decreasing multi-step function
\begin{equation}\label{g_x}
g(x)= \sum_{i=0}^{M-1} \gamma_{i}\sigma(x-\theta_{i}),
\end{equation}
where $M$ is the number of levels and $\sigma(.)$ is the Heaviside step function. Letting $\beta_i=\sum_{n=0}^i\gamma_n$ we have $\theta_{i+1}$ as the
value of $x$ at which $g(x)$ jumps from value $\beta_i$ to
$\beta_{i+1}$. Since we assume
$x_{\min}=\theta_{0}<\theta_{1}<\theta_{2}<...<\theta_{M-1}<\theta_M=x_{\max}$, the
optimal  $g(x)$ is unique. This latter requirement means that we consider only the case of monotonically non-decreasing
(sigmoidal) tuning curves. Without this restriction it is not possible to find a unique solution and hence this study does not generalize to non-monotonic tuning curves. This is not highly restrictive, since sigmoidal tuning curves are widely observed in many sensory modalities~\cite{Salinas.06}. The mutual information of the neural
channel can be written as
\begin{equation}
\label{m4_information}
I(x;k)=\sum_{k=0}^{\infty}\sum_{i=0}^{M-1}\alpha_{i} Q[k|\beta_{i}] \log_{2}
\frac{Q[k|\beta_{i}]}{\sum_{n=0}^{M-1}\alpha_{n} Q[k|\beta_{n}]},
\end{equation}
where $\alpha_{i}=\int_{\theta_{i}}^{\theta_{i+1}}dx P_{x}(x)$. The optimal function $g(x)$ cannot be easily found in an
analytical form using variational principles, because it leads to a set of
transcendental equations.  Therefore we use stochastic gradient descent methods to solve for the optimal $P_\nu(\nu)$. For alternative methods see e.g.~\cite{McDonnell.09PRE}.

\begin{figure}[t]
\begin{center}
\includegraphics[width=0.85\textwidth]{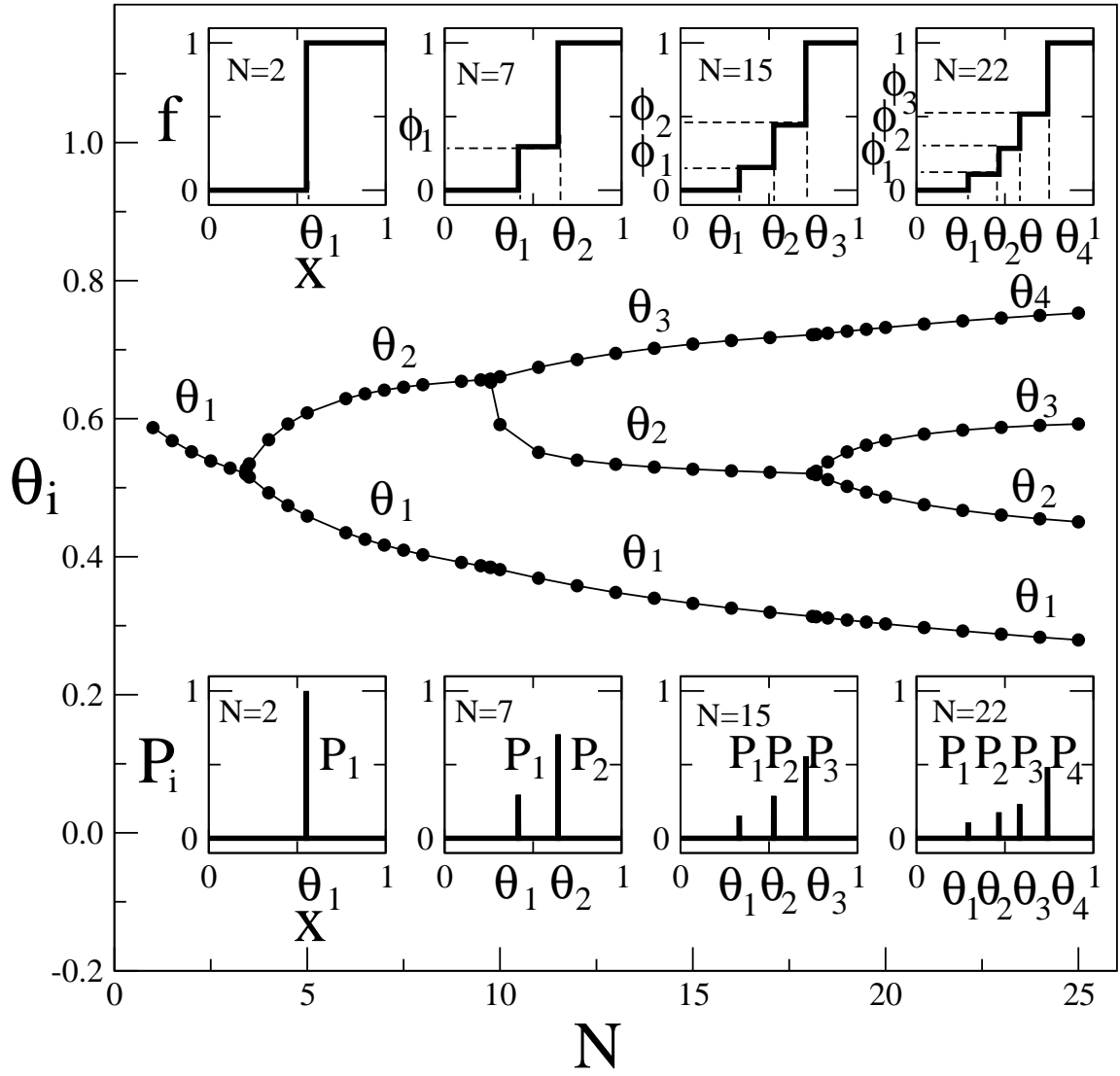}
\caption{The partition boundaries $\theta_{i}$ against the maximum
mean spike count, $N=T\nu_{\max}$. Also shown for
$N=2,7,15$ and $22$, are the optimal $f(x)$ (top insets) and the population distributions (bottom insets). The parameters are $x_{\min}=0$, $x_{\max}=1$, $\nu_{\min}=0$, and $x$ uniformly distributed.}\label{figure1}
\end{center}
\end{figure}

\begin{figure}[t]
\begin{center}
\includegraphics[width=0.85\textwidth]{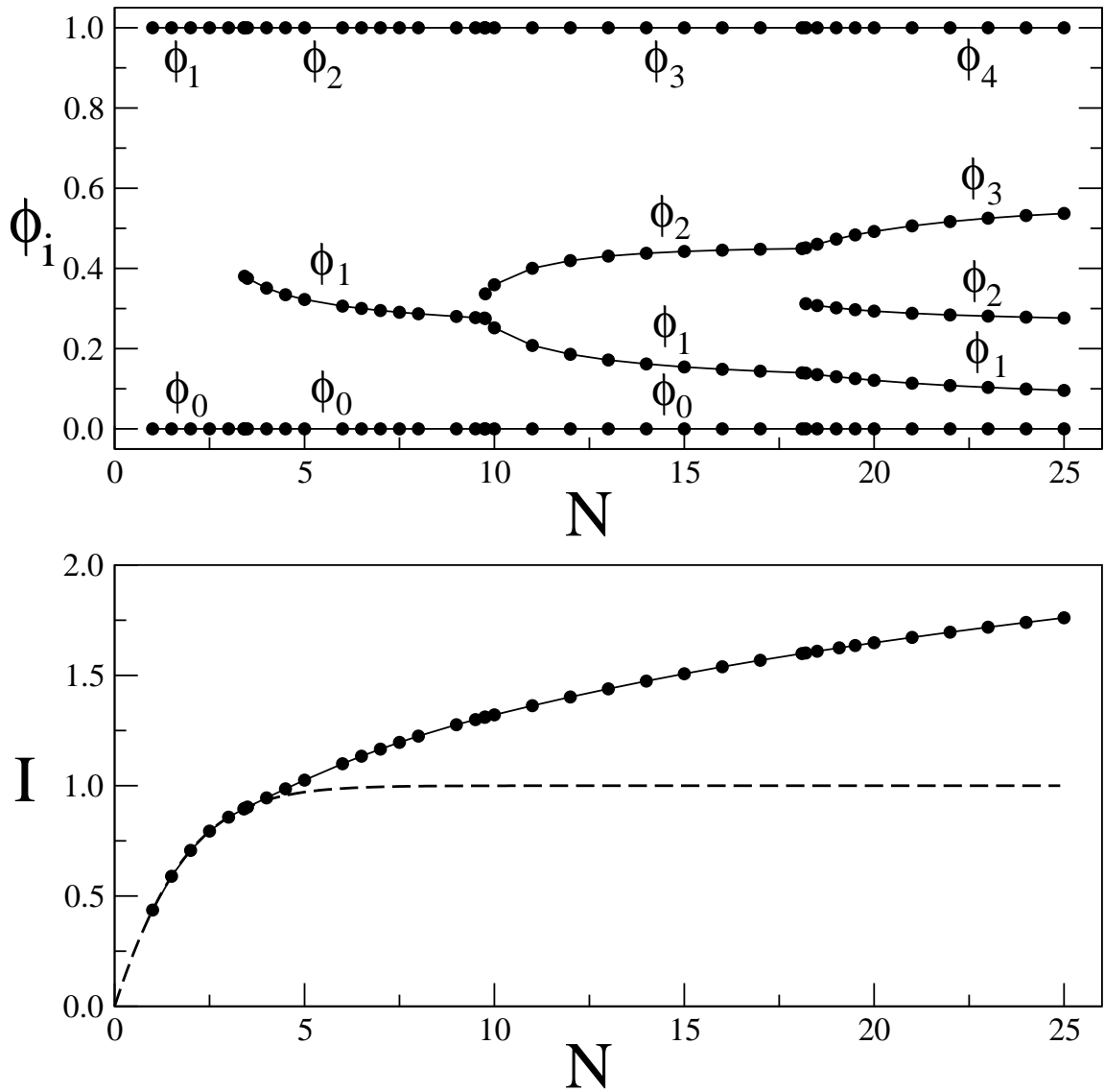}
\caption{The set of optimal firing rates, $\phi_i$, $i=0,..,M-1$, for the overall population with $M-1$ subpopulations, as a function of $N$ (top), and the mutual information for the optimal solution (bottom).
The dashed line shows the mutual information that would result if binary
coding were utilized~\cite{Shamai90}.}\label{figure2}
\end{center}
\end{figure}

Fig.~\ref{figure1} shows the main results of our study. The upper insets display the normalized optimal tuning curve,
$f(x)\equiv g(x)/N$, for four different values of maximum mean spike count,
$N$.  Fig.~\ref{figure2} shows the overall population normalized
firing rates, $\phi_i\equiv\beta_{i}/N$, as well as the mutual information corresponding to
the optimal solution. Note that $I(x;k)$ in Eq.~(\ref{m4_information}) is parameterized entirely by the set $\alpha_i,~\beta_i~,i=0,..,M-1$, and it is these parameters that are optimized. The set of $\theta_i$--s required for the optimal $g(x)$ can be obtained for any given $P_x(x)$ from the $\alpha_i$. Hence, in Fig~\ref{figure1}, without loss of generality we have assumed that the stimulus is uniformly distributed on $[0,1]$. Similarly, the $\gamma_i$--s follow from $\beta_i$.

For small $N<3$, only two firing rates are observed; for values of
$x<\theta_1$, $f(x)=0$ (the absence of  firing) while for larger values of $x$,
$f(x)=1$ (firing at the maximum allowable spike rate). This form of optimal binary coding has been predicted previously
for Poisson neurons using estimation theory~\cite{Bethge1,Bethge2}. It also agrees with the well known result that a binary source maximizes information through a Poisson channel when the input can switch instantaneously between states~\cite{Kabanov,Johnson.08}.

As $N$ is increased, the number of steps in the optimal tuning curve
increases; e.g. for $N=7$, two steps are observed giving rise to a ternary coding scheme, for $N=15$ three steps are observed giving a 4-ary
(quaternary) coding. In general, an $M$-ary code
will be optimal with increasing $N$. As $N\rightarrow\infty$ we predict that the optimal tuning curve will converge to a continuous function~\cite{Brunel.98,McDonnell.PRL08}. Fig.~\ref{figure1}
shows how the partition boundaries, $\theta_i$, vary as $N$ is increased; new boundaries can be seen to emerge via
phase transitions. These appear to be continuous and hence are akin to second order phase
transitions of the optimal tuning curve.

Our findings of an optimal $M$-ary code are in agreement with isomorphic results on the information maximizing source distribution for Poisson direct detection photon channels with imposed bandwidth constraints~\cite{Shamai90}. In our context, a bandwidth constraint is equivalent to allowing $N>1$. We further note that the bifurcation structure in Fig.~\ref{figure1} is qualitatively similar to information optimization results in~\cite{Rose,McDonnell} for systems that are quite different to Poisson neurons.

One way of interpreting our results is that the steps in the optimal $f(x)$ partition the stimulus into regions associated with neural {\em subpopulations}. For example, suppose an overall population consists of $K$ neurons and $M-1$ sub-populations, within which each neuron is identical, and binary with rates $0$ and $\gamma_i$. Since the neurons are Poisson, the sum of the $K$ individual normalized firing rates is equal to $f(x)$. For overall binary coding, the only way of achieving
$f(x)$ would be a single sub-population, where each neuron is identical, and able to fire at two rates, $\phi_0/K=0$ and $\phi_1/K=\frac{1}{K}$, where rate $\phi_1/K$ is activated when $x\ge\theta_1$. For the ternary case, there would be two subpopulations, of sizes $J$ and $K-J$, with individual normalized firing rates $\phi_1/J$ and $(1-\phi_1)/(K-J)$, so that the overall population has 3 rates: $0,~\phi_1$ and $1$, as shown in Fig.~\ref{figure2}. The first subpopulation would only be activated when $x>\theta_1$ and the second when $x>\theta_2$.

We can estimate the sizes of the subpopulations in our example as follows.
Since the sizes of the subpopulations are proportional to the integrated
firing rates, the neurons for ternary coding are distributed with
probabilities $P_1=\phi_{1}=\gamma_1/N$ and $P_2=1-\phi_{1}$ respectively.
The quaternary coding scheme for $N=15$ has three subpopulations with
optimal individual firing rates proportional to $\gamma_1,~\gamma_2$ and $1-\gamma_1-\gamma_2$,
and overall rates  $0,~\phi_1,~\phi_2$ and $1$. The sizes of the subpopulations
are therefore $P_1\propto\gamma_1=N\phi_1$, $P_2\propto\gamma_2=N(\phi_2-\phi_1)$ and
$P_3=1-P_1-P_2$, as shown in the lower insets in Fig.~\ref{figure1}.

 Our results lead to two main predictions for information-optimal neural coding: (i) that the tuning curves are discrete; and (ii) that neural populations should form subpopulations.

The prediction of subpopulations seems to have some correspondence with known results and we take, as an example, coding of sound level (at a fixed frequency) by the auditory system. Inner hair cells (the sensory receptors that transduce sounds into neural activity) are each connected to approximately 15 separate afferent nerve fibers~\cite{Spoendlin.89}.  Physiological studies suggest that these fibers can be grouped into two or three subpopulations based on their threshold to sound level~\cite{Liberman82,Jackson.05}.

The presence of three subpopulations would suggest a quaternary code ($M = 4$) is used. From Fig.~\ref{figure2}, this is optimal when $N\sim 15$. If each afferent fires at a rate of $\sim 100$ spikes/s then the time required for the population to generate 15 action potentials is $\sim 10$~ms. This timescale agrees with the classical temporal-window model proposed to explain the human auditory system's temporal resolution, in which temporal integration is performed by a sliding window with an equivalent rectangular duration of $\sim7-13$ ms~\cite{Plack.90}. Hence, quaternary coding is consistent with the known parameters of auditory coding and perception.

We now turn our attention to the prediction that optimal-information tuning curves should be discrete. This certainly seems to be inconsistent with physiologically measured tuning curves. However, evidence for binary tuning curves does exist---for a discussion see~\cite{Bethge1}. There may well be other reasons why they are not commonly observed in practice. For example, the steepness of the slope of the sigmoid is known to depend strongly on the nature of the signal~\cite{sachs.74} and measurement window~\cite{meddis.88}.  Consequently, measuring the tuning curve with the `right' stimulus (feature), $x$, and window duration may be crucial to observing a rapidly increasing curve that approximates a step increase.  At present it is not clear if such experiments have been performed.

Of course, it is also possible that neural sensory systems are not optimized for the transmission of information, although we note that minimization of MSE also leads to the prediction of binary tuning curves for short decoding windows~\cite{Bethge1}. Alternatively, the model we consider may need revision to make it physiologically more realistic. For example, it may be necessary to include other sources of noise and to take into account non-Poisson statistics. Realistic signal statistics will also need to be considered. Other constraints, such as metabolic penalty, may also be important, although this is not likely to change the conclusion that information-optimal tuning curves are discrete~\cite{Huang.05}. We hope our results will inform future discussion on this topic and motivate further studies.

\begin{acknowledgments}
We thank Steven Holmes, Riccardo Mannella and Jan A. Freund for fruitful
discussions. This work was funded by the EPSRC
(grant GR/R35650/01 and EP/D05/1894/1(P)). Mark D. McDonnell is funded by the Australian Research
Council, Grant No. DP0770747.
\end{acknowledgments}

\clearpage

\end{document}